

\font\rmu=cmr10 scaled\magstephalf
\font\bfu=cmbx10 scaled\magstephalf

\font\it=cmti10 scaled \magstephalf

\rmu

\font\rmus=cmr8
\font\rmuss=cmr6
\font\mait=cmmi10 scaled\magstephalf
\font\maits=cmmi7 scaled\magstephalf
\font\maitss=cmmi7
\font\msyb=cmsy10 scaled\magstephalf
\font\msybs=cmsy8 scaled\magstephalf
\font\msybss=cmsy7
\font\bfus=cmbx7 scaled\magstephalf
\font\bfuss=cmbx7
\font\cmeq=cmex10 scaled\magstephalf

\textfont0=\rmu
\scriptfont0=\rmus
\scriptscriptfont0=\rmuss

\textfont1=\mait
\scriptfont1=\maits
\scriptscriptfont1=\maitss

\textfont2=\msyb
\scriptfont2=\msybs
\scriptscriptfont2=\msybss

\textfont3=\cmeq
\scriptfont3=\cmeq
\scriptscriptfont3=\cmeq

\newfam\bmufam  \textfont\bmufam=\bfu
      \scriptfont\bmufam=\bfus \scriptscriptfont\bmufam=\bfuss

\hsize=15.5cm
\vsize=22cm
\baselineskip=16pt   
\parskip=16pt plus  2pt minus 2pt

\def\a{\alpha}
\def\b{\beta}

\def\l{\lambda}

\def\semi{\bigcirc\kern-1em{s}\;}

\def\ni{\noindent}

\def\R{{\rm I\!R}}

\def\Q{{\mathchoice
{\setbox0=\hbox{$\displaystyle\rm Q$}\hbox{\raise 0.15\ht0\hbox to0pt
{\kern0.4\wd0\vrule height0.8\ht0\hss}\box0}}
{\setbox0=\hbox{$\textstyle\rm Q$}\hbox{\raise 0.15\ht0\hbox to0pt
{\kern0.4\wd0\vrule height0.8\ht0\hss}\box0}}
{\setbox0=\hbox{$\scriptstyle\rm Q$}\hbox{\raise 0.15\ht0\hbox to0pt
{\kern0.4\wd0\vrule height0.7\ht0\hss}\box0}}
{\setbox0=\hbox{$\scriptscriptstyle\rm Q$}\hbox{\raise 0.15\ht0\hbox to0pt
{\kern0.4\wd0\vrule height0.7\ht0\hss}\box0}}}}
\def\C{{\mathchoice
{\setbox0=\hbox{$\displaystyle\rm C$}\hbox{\hbox to0pt
{\kern0.4\wd0\vrule height0.9\ht0\hss}\box0}}
{\setbox0=\hbox{$\textstyle\rm C$}\hbox{\hbox to0pt
{\kern0.4\wd0\vrule height0.9\ht0\hss}\box0}}
{\setbox0=\hbox{$\scriptstyle\rm C$}\hbox{\hbox to0pt
{\kern0.4\wd0\vrule height0.9\ht0\hss}\box0}}
{\setbox0=\hbox{$\scriptscriptstyle\rm C$}\hbox{\hbox to0pt
{\kern0.4\wd0\vrule height0.9\ht0\hss}\box0}}}}

\font\fivesans=cmss10 at 4.61pt
\font\sevensans=cmss10 at 6.81pt
\font\tensans=cmss10
\newfam\sansfam
\textfont\sansfam=\tensans\scriptfont\sansfam=\sevensans\scriptscriptfont
\sansfam=\fivesans
\def\sans{\fam\sansfam\tensans}
\def\Z{{\mathchoice
{\hbox{$\sans\textstyle Z\kern-0.4em Z$}}
{\hbox{$\sans\textstyle Z\kern-0.4em Z$}}
{\hbox{$\sans\scriptstyle Z\kern-0.3em Z$}}
{\hbox{$\sans\scriptscriptstyle Z\kern-0.2em Z$}}}}

\newcount\foot
\foot=1
\def\note#1{\footnote{${}^{\number\foot}$}{\ftn #1}\advance\foot by 1}

\def\frac#1#2{{#1\over #2}}
\def\text#1{\quad{\hbox{#1}}\quad}

\font\ch=cmbx12 scaled\magstephalf
\font\ftn=cmr8 scaled\magstephalf

\font\it=cmti10 scaled\magstephalf

\font\titch=cmbx12 scaled\magstep2
\font\titname=cmr10 scaled\magstep2
\font\titit=cmti10 scaled\magstep1
\font\titbf=cmbx10 scaled\magstep2

\nopagenumbers


\line{\hfil SU-GP-93/3-2}
\line{\hfil March 8, 1993}
\vskip3cm
\centerline{\titch LOOP VARIABLE INEQUALITIES IN}
\vskip.5cm
\centerline{\titch GRAVITY AND GAUGE THEORY}
\vskip2cm
\centerline{\titname R. Loll }
\vskip.5cm
\centerline{\titit Physics Department, Syracuse University,}
\vskip.2cm
\centerline{\titit Syracuse, NY 13244,}
\vskip.2cm
\centerline{\titit U.S.A.}

\vskip4cm
\centerline{\titbf Abstract.\hfill}
\ni
We point out an incompleteness of formulations of
gravitational and gauge theories that use
traces of holonomies around closed curves as their basic variables. It
is shown that in general
such loop variables have to satisfy certain inequalities if
they are to give a description equivalent to the usual one in terms of
local gauge potentials.

\vfill\eject
\footline={\hss\tenrm\folio\hss}
\pageno=1

\line{\ch  1 Introduction\hfil}

Since the introduction of an $SL(2,\C)$-Yang-Mills connection $A$ as the basic
configuration variable for pure gravity by Ashtekar [1,2], the idea of
using Wilson loops, i.e. non-local, gauge-invariant quantities
depending on the holonomy

$$
U_A(\a):=\, {\rm P}\,\exp \oint_\a A_\mu d\a^\mu \eqno(1.1)
$$

\ni around closed curves $\a$ in the space(-time) manifold $\Sigma$,
has found wide application. (The symbol P in (1.1) indicates path ordering
along $\a$.) The (normalized) Wilson loop, or traced holonomy,
of $\a$ is the quantity

$$
T_A(\a):= \frac{1}{N}\,{\rm Tr} \,U_A(\a),\eqno(1.2)
$$

\ni with $N$ denoting the dimension of the matrix representation.
It is a main ingredient in attempts to quantize
the theory non-perturbatively [3] (see also [4] for a recent review).
One motivation for adopting a ``pure loop approach" is the idea that,
although classically equivalent, it may eventually lead to an inequivalent
quantum representation, whose non-perturbative features do not resemble
those of the quantum connection representation. However,
if one wants to reformulate gravity (or
non-abelian gauge
theory) in terms of these gauge-invariant loop variables, such that the
original
(gauge-{\it co}variant) potentials $A$ do not any more appear
in the description, one
has to make sure that the two formulations are indeed equivalent.
This means that, at the kinematical level, there should be a one-to-one
correspondence between the gauge potentials $A(x)$ modulo local gauge
transformations and the loop variable configurations. In other words, one
has to impose appropriate conditions on the set of complex-valued,
``bare" loop functions,

$$
\{ T(\a)\, |\, \a\;\, {\rm a\; closed\; curve\; in}\; \Sigma \}, \eqno(1.3)
$$

\ni in order to ensure they are the traces of holonomies of a given gauge
group $G$. Finding a complete set of such conditions involves a number of
subtleties, and for general gauge group $G$ is an unsolved problem.

On the one hand, there are the so-called Mandelstam constraints, certain
$G$-dependent,
algebraic constraints among the loop variables, which have
been discussed by several authors [5,6,7,8].
The Mandelstam constraints for $SL(2,\C)$
(and for any of its subgroups) are given by

$$
T(\a)T(\b)-\frac12 (\,T(\a\circ_x\b)+T(\a\circ_x\b^{-1})\,)=0.\eqno(1.4)
$$

\ni The notation in equation (1.4)
refers to a configuration of two loops $\a$ and $\b$ intersecting in a
point $x$, the base point for all loops, with $\circ_x$ denoting the loop
composition in $x$.
Polynomial constraints of this kind lead to an overcompleteness of the
set of loop functions. There are several possibilities of
implementing them either classically or in the quantum theory. A maximal
identification
of the independent physical degrees of freedom seems to be possible only
in the regularized lattice version of the theory [9]. An
algebraic treatment of the Mandelstam constraints for $SL(2,\C)$
has been given by Ashtekar and Isham [10], who construct an algebra of
$T$-variables in such a way that the constraints (1.4)
are satisfied automatically
and continue to hold in any quantum representation of that algebra.

On the other
hand, there is the problem of selecting an appropriate set of closed curves
on which the loop variables are to be defined. It turns out that the
fundamental objects are not the loops themselves, but certain equivalence
classes of loops (under reparametrization, retracing, orientation reversal
etc., see [11] for a detailed discussion and references).
Since the distinction is not essential in the present context, I will
continue to use the word `loop' for such an equivalence class.

Moreover, there are examples of gauge groups where the traced holonomies
do not contain all the local gauge-invariant information. For example,
there are  non-trivial subgroups of ``null rotations" of $SL(2,\C)$
which are mapped into a single
configuration in terms of the variables $T(\a)$ [12].

The purpose of this paper is to point out that, in addition to the
well-known Mandelstam constraints, certain inequalities have to be satisfied
by the loop variables $T(\a)$ in order to achieve equivalence with the
usual connection representation. These inequalities cannot be derived from
the Mandelstam constraints. This renders quantization approaches based on
loop variable algebras, and which are constructed independently of
quantum representations based on the connection $A$, incomplete.

Because of its relevance for gravitational theories.
I will discuss the case of gauge group $G=SL(2,\C)$ and two of its
subgroups ($SU(2)$ and $SU(1,1)$),
in the defining representation by $2\times 2$ complex matrices.
However, similar problems are expected to occur
for other non-abelian gauge groups and representations. - The next section
contains the derivation of some specific examples of inequalities
between the loop variables. The consequences of the present result for general
(quantum) loop approaches are outlined in Sec.3.

\vskip1.5cm

\line{\ch 2 Loop inequalities for subgroups of $SL(2,\C)$ \hfil}

The ``gauge group" for the gravitational field in Ashtekar's Hamiltonian
reformulation
is the non-compact Lie group $SL(2,\C)$. However, the physical configurations
take their values in a particular $SO(3)$-subgroup of $SL(2,\C)$,
which is projected out by a set of
reality conditions on phase space.
Unfortunately, no manageable form for these reality
conditions is known in the loop formulation, where the basic variables
are given by the Wilson loops (1.2).

In the following I will look at two other subgroups of $SL(2,\C)$, one of
type $SU(2)$, and the other of type $SU(1,1)$, which are embedded into
$SL(2,\C)$ in a simple way. The former is of course of
interest for Yang-Mills theory, whereas the latter appears in reduced
gravitational models such as 2+1 gravity (see [13] for a treatment
in the loop formulation), and
3+1 gravity with one (space-like) Killing vector field.

The new inequalities that have to be satisfied by the loop variables describing
these subgroups become apparent only when one uses the original explicit
representation of the $2\times 2$ holonomy matrices. From these one derives
inequalities between the group parameters, which then
translate into gauge-invariant inequalities among the loop variables.

We will first treat the case of $SU(2)\subset SL(2,\C)$, given by the
subgroup of matrices of the form

$$
U_{\a}=\left(\matrix{\;\a_1+i\a_2&\a_3+i\a_4\cr -\a_3+i\a_4&\a_1-i\a_2\cr}
\right),\;
U_{\b}=\left(\matrix{\;\b_1+i\b_2&\b_3+i\b_4\cr -\b_3+i\b_4&\b_1-i\b_2\cr}
\right),\; {\rm etc.}, \eqno(2.1)
$$

\ni with real parameters $\a_i$, $\b_i$, ... , subject to the constraints
$\sum_i \a_i^2=1$, $\sum_i \b_i^2=1$, etc.. The indices $\a$ and $\b$ on the
holonomy matrices $U$ label loops starting and
ending at a given base point $x$. In this parametrization, the group variables
$\a_i$ can be thought of as the components of a unit vector $\vec\a$ imbedded
into $\R^4$. The relevant traced
holonomies are

$$
\eqalign{T(\a)=\,&\a_1\cr
T(\a\circ_x \b)=&\,\a_1\b_1 -\a_2\b_2-\a_3\b_3-\a_4\b_4=:
\a_1\b_1 -\vec\a_{\perp}\cdot\vec \b_{\perp},\cr}\eqno(2.2)
$$

\ni where a convenient vectorial notation has been introduced for the 2-,
3- and 4-components of the vectors $\vec\a$ and $\vec\b$. The conditions
$\sum_i \a_i^2=1$ imply the well-known fact that the loop variables $T(\a)$
for $G=SU(2)$ are bounded functions on loop space, namely,

$$
-1\leq T(\a) \leq 1,\qquad \forall\a.\eqno(2.3)
$$

\ni This boundedness was used in a crucial way by Ashtekar and Isham in their
application of Gel'fand spectral theory to the abelian
algebra of $T$-variables. However, there
exist more complicated inequalities, which so far seem to have been overlooked.
For the case of $SU(2)$, they are
easily derived from geometric considerations, but in view of less
straightforward cases (such as the one of $SU(1,1)$ presented below),
I will sketch their proper mathematical derivation.

I will show that, given arbitrary values for the traced holonomies around
two loops $\a$ and $\b$, $T(\a)=c$, $T(\b)=c'$, say, with $c,c'\in [-1,1]$,
the Wilson loop
$T(\a\circ_x \b)$ cannot assume arbitrary values in the interval $[-1,1]$, but
rather has to obey an inequality, depending on $c$ and $c'$. For this purpose,
it is convenient (but not necessary) to consider linear combinations of the
form

$$
L_2(\a,\b):=\frac12 (\,T(\a\circ_x\b^{-1})-T(\a\circ_x\b)\,)
            =\vec \a_{\perp}\cdot\vec \b_{\perp},\eqno(2.4)
$$

\ni which were first introduced in [14].
(Note that the {\it sum} of the two terms appearing on the right-hand
side of the definition, $T(\a\circ_x\b^{-1})+T(\a\circ_x\b)$,
by virtue of (1.4) is an algebraically
dependent quantity.) In order to find the (conditional) extrema of
$L_2(\a,\b)$, for fixed $T(\a)$ and $T(\b)$, one best uses the method of
Lagrangian multipliers. Extremal points in $\R^8$ (with coordinates
$\a_i$, $\b_i$, $i=1,\dots,4$) must satisfy

$$
\nabla L_2(\a,\b)-\l\nabla (\a_1 -c) -\l'\nabla (\b_1 -c') -\mu\nabla
(\sum_i \a_i^2\; -1) -\mu'\nabla(\sum_i \b_i^2\; -1)=0,\eqno(2.5)
$$

\ni where $\l$, $\l'$, $\mu$ and $\mu'$ are real Lagrange multipliers.
{}From (2.5) one derives the condition $\vec\a_\perp = k \vec\b_\perp$,
$k\in\R\backslash\{ 0\}$, for local extrema.
To determine the nature of these extrema, one
has to compute the Hessian in these points (in a suitable local basis of
tangent vectors), and its signature. This computation yields an
absolute maximum in points with $\vec\a_\perp = k \vec\b_\perp$,
$k = \frac{\sqrt{ 1-\a_1^2}}{\sqrt{ 1 - \b_1^2}}$, and an absolute
minimum in
points with $\vec\a_\perp = k \vec\b_\perp$,
$k = - \frac{\sqrt{ 1-\a_1^2}}{\sqrt{ 1 - \b_1^2}}$. Rephrased
in terms of loop variables, the result reads as follows:
{\it for the $SU(2)$-subgroup
given by matrices of the form (2.1),
and for given $T(\a)$ and $T(\b)$, $L_2(\a,\b)$ may only
assume values in the interval}

$$
-\sqrt{(1- T(\a)^2)(1- T(\b)^2)}\leq L_2(\a,\b)\leq
                \sqrt{(1- T(\a)^2)(1- T(\b)^2)}.\eqno(2.6)
$$

\ni The equivalent statement in terms of the original traced holonomies
is obtained by substituting $L_2$ according to (2.4).

Let us now turn to the case of $SU(1,1)\subset SL(2,\C)$, given by the
subgroup of matrices of the form

$$
U_{\a}=\left(\matrix{\;\a_1+i\a_2&\a_3+i\a_4\cr \a_3-i\a_4&\a_1-i\a_2\cr}
\right),\;
U_{\b}=\left(\matrix{\;\b_1+i\b_2&\b_3+i\b_4\cr \b_3-i\b_4&\b_1-i\b_2\cr}
\right),\; {\rm etc.}, \eqno(2.7)
$$

\ni again with real parameters $\a_i$, $\b_i$, ... , but now subject to the
constraints $\a_1^2 +\a_2^2 -\a_3^2 -\a_4^2 =1$,
$\b_1^2 +\b_2^2 -\b_3^2 -\b_4^2 =1$, etc.. As a consequence of the
non-compactness of $SU(1,1)$, the loop variables $T(\a)$ are no longer
bounded functions on loop space. Adopting the same definition for the
variables $L_2$ as in (2.4) above leads to

$$
L_2(\a,\b)=\a_2\b_2 -\a_3\b_3 -\a_4\b_4. \eqno(2.8)
$$

\ni Note that, as in the case of $SU(2)$, all the traced holonomies $T(\a)$
are {\it real} functions on loop space.
Following the same strategy as for $SU(2)$ leads to the condition

$$
\nabla L_2(\a,\b)-\l\nabla (\a_1 -c) -\l'\nabla (\b_1 -c') -\mu\nabla
(\a_1^2 +\a_2^2 -a_3^2 -a_4^2 -1) -\mu'\nabla
(\b_1^2 +\b_2^2 -\b_3^2 -\b_4^2 -1)=0\eqno(2.9)
$$

\ni for local extrema. Again this yields as solutions points with
$\vec\a_\perp = k \vec\b_\perp$, $k\in\R\backslash\{ 0\}$.
The computation of the Hessian
is slightly more involved, and one finds that its signature is not necessarily
positive or negative (semi-)definite. The final result is as follows:
$L_2(\a,\b)$ has a local maximum in points for which
$\vec\a_\perp = k \vec\b_\perp$, $k =- \frac{\sqrt{ 1-\a_1^2}}
{\sqrt{ 1 - \b_1^2}}<0$, and $1-T(\a)^2\geq 0$
($\iff 1-T(\b)^2\geq 0$), and a local minimum in points with
$\vec\a_\perp = k \vec\b_\perp$,
$k = \frac{\sqrt{ 1-\a_1^2}}{\sqrt{ 1 - \b_1^2}}>0$, and $1-T(\a)^2\geq 0$
($\iff 1-T(\b)^2\geq 0$). All other extrema are saddle points.
Again this result may be restated in a gauge-invariant way in terms of the
loop variables only: {\it for the $SU(1,1)$-subgroup
given by matrices of the form (2.7),
and for given $T(\a)$ and $T(\b)$, {\it and} if both
$1-T(\a)^2\geq 0$ and $1-T(\b)^2\geq 0$,
$L_2(\a,\b)$ may only assume values}

$$
L_2(\a,\b)\leq -\sqrt{(1- T(\a)^2)(1- T(\b)^2)} \quad {\rm or}\quad
L_2(\a,\b)\geq \sqrt{(1- T(\a)^2)(1- T(\b)^2)}.\eqno(2.10)
$$

\ni {\it For other values of $T(\a)$ and $T(\b)$ there are no restrictions, and
$L_2(\a,\b)$ may assume any value on the real line.}

\vskip1.5cm

\line{\ch 3 Conclusions\hfil}

In the preceding section we derived some examples of
inequalities that have to be
satisfied by loop functions $T(\a)$ in order that they correspond to the
traced holonomies of particular subgroups of $SL(2,\C)$. These inequalities
are yet another manifestation of the non-linearities of the underlying
physical configuration spaces.
Incidentally, our derivation
provides an answer to a question posed in [10], namely, how to distinguish
between the subgroups $SU(2)$ and $SU(1,1)$ of $SL(2,\C)$, without making
reference to the gauge potentials $A$.

The question remains of how such inequalities, whenever they appear,
can be incorporated in the
loop formulation of gravitational and gauge theories.
It can easily be shown that inequalities are present even if one
considers only loop configurations without (self-) intersections.
This means that such configurations
do not just represent isolated points in the configuration space,
but are abundant. Furthermore, inequalities such as (2.6) and (2.10) do
not exhaust the set of all restrictions for a given gauge group, as is
exemplified by the loop treatment of the $1\times 1$-lattice gauge theory
[8]. They depend in an essential way on the geometry of the loop
configuration, and there is no obvious
systematic way of identifying all of them,
unless one works in a lattice regularization where the number of loop
variables can be greatly reduced.

This observation is relevant for 3+1-dimensional gravity insofar as also there
the physical data take values in a subalgebra of $sl(2,\C)$, namely, in
$so(3)$. The embedding of the subalgebra is much more complicated in this case,
since it is defined in terms of non-linear equations not on the configuration,
but on the {\it phase} space of the canonical variable pairs
$(A,E)$.
The examples treated above suggest that
even if we were to find a way to implement this restriction to $so(3)$
in the loop representation (for example, by finding a suitable form of
the reality conditions in terms of loop variables), we may still have to deal
with additional inequalities of the kind described in Sec.2
on the thus reduced space of loop variables.

Inequalities are notoriously hard to implement in canonical quantization
procedures, and their importance for gravity in the metric representation
(where one requires det$\,g_{ij} >0$)
has repeatedly been emphasized by Isham (see, for example, [15]).
A similar problem may appear in the loop formulation of 3+1 gravity, unless
one wants to rely on a loop transform of quantum states
(along the lines discussed in [10]) from the
connection representation to the loop representation, whose existence
however has not been established. The presence of
loop variable inequalities is directly relevant for pure loop
formulations of Yang-Mills theory and various reduced gravitational models.
This observation may not come entirely as a surprise, since
in the case of Yang-Mills theory it is well-known
that inequalities appear in certain non-standard parametrizations
of the theory (see, for example, the flux representation discussed in [16]). -
Unless the existence of such loop variable
inequalities is taken properly into account, any formulation of gravitational
and gauge theories based exclusively on the Wilson loops (1.2) remains
inequivalent to its usual formulation in terms of connection variables.

\ni{\it Acknowledgement.} I wish to thank the members of the Syracuse
Relativity Group for their comments during the Monday relativity tea
where these results were first presented.
\vfill\eject

\line{\ch References\hfil}
\vskip0.8cm

\item{[1]} Ashtekar, A.: New variables for classical and quantum gravity,
  Phys. Rev. Lett. 57 (1986) 2244-2247
\item{[2]} Ashtekar, A.: New Hamiltonian formulation of general relativity,
  Phys. Rev. D36 (1987) 1587-1602
\item{[3]} Rovelli, C. and Smolin, L.: Loop space representation of
  quantum general relativity, Nucl. Phys. B331 (1990) 80-152
\item{[4]} Rovelli, C.: Ashtekar formulation of general relativity and
  loop-space non-perturbative quantum gravity: A report, Class. Quantum
  Grav. (1991) 1613-1675
\item{[5]} Gliozzi, F. and Virasoro, M.A.: The interaction among dual
  strings as a manifestation of the gauge group, Nucl. Phys. B164 (1980)
  141-151
\item{[6]} Giles, R.: Reconstruction of gauge potentials from Wilson loops,
  Phys. Rev. D24 (1981) 2160-2168
\item{[7]} Gambini, R. and Trias, A.: Gauge dynamics in the
  C-representation, Nucl. Phys. B278 (1986) 436-448
\item{[8]} Loll, R.: Yang-Mills theory without Mandelstam constraints,
  to appear in Nucl. Phys.
\item{[9]} Loll, R.: Independent SU(2)-loop variables and the reduced
  configuration space of SU(2)-lattice gauge theory, Nucl. Phys.
  B368 (1992) 121-142
\item{[10]} Ashtekar, A. and Isham, C.J.: Representations of the
  holonomy algebras of gravity and non-Abelian gauge theories, Class.
  Quant. Grav. 9 (1992) 1433-1467
\item{[11]} Loll, R.: Loop approaches to gauge field theory, in: Memorial
  Volume for M.K. Polivanov, Teor. Mat. Fiz. 91 (1992)
\item{[12]} Goldberg, J.N., Lewandowski, J. and Stornaiolo, C.:
  Degeneracy in loop variables, Comm. Math. Phys. 148 (1992) 377-402
\item{[13]} Marolf, D.M.: Loop representations for 2+1 gravity on a torus,
  Syracuse preprint SU-GP-93/3-1
\item{[14]} Loll, R.: A new quantum representation for canonical gravity and
  SU(2) Yang-Mills theory, Nucl. Phys. B350 (1991) 831-860
\item{[15]} Isham, C.J.: Topological and global aspects of quantum theory,
  {\it in} Relativity, groups and topology II, ed.
  B.S. DeWitt and R. Stora (North-Holland, Amsterdam, 1984) 1059-1290
\item{[16]} Furmanski, W. and Kolawa, A.: Yang-Mills vacuum: An attempt
  at lattice loop calculus, Nucl. Phys. B291 (1987) 594-628

\end